\documentclass[final,5p,times,twocolumn,numbers]{elsarticle}

\usepackage{amssymb}
\usepackage{lipsum}

\usepackage{graphicx}  
\usepackage{subcaption}
\usepackage{dcolumn}   
\usepackage{bm}         
\usepackage{amssymb}   
\usepackage{amsmath}
\usepackage{mathrsfs}  
\usepackage{graphicx,color} 
\usepackage[english]{babel}     
\usepackage{graphicx,wrapfig,lipsum}      
\usepackage{multimedia}    
\usepackage[mathscr]{eucal}  
\usepackage{bm}              
\usepackage{amssymb}        
\usepackage{amsmath}     
\usepackage{mathrsfs}  
\usepackage[mathscr]{eucal}   

\usepackage{showlabels}
\newcommand{\be}{\begin{equation}}
	\newcommand{\ee}{\end{equation}}
\newcommand{\bea}{\begin{align}}
	\newcommand{\eea}{\end{align}}

\makeatletter
\providecommand*{\lambdabar}{{\mathpalette\@TM@lambdabar\relax}}
\def\@TM@lambdabar#1#2{%
	\setbox0=\hbox{\m@th$#1\lambda$}%
	\setbox1=\hb@xt@\z@{\m@th\hss$#1\mathchar'26$}%
	\box0\lower.15\ht1\box1
} 
\def\ket{\@ifstar\@ket\@@ket}
\newcommand*{\@ket}[1]{\mkern2mu \lvert #1 \rangle\!}
\newcommand*{\@@ket}[1]{\mkern2mu \lvert #1 \rangle}
\def\bra{\@ifstar\@bra\@@bra}
\newcommand*{\@bra}[1]{\!\langle #1 \rvert \mkern2mu}
\newcommand*{\@@bra}[1]{\langle #1 \rvert \mkern2mu}
\makeatother 
\bmdefine\va{a}
\bmdefine\vb{b}
\bmdefine\vc{c}
\bmdefine\ve{e}
\bmdefine\vf{f}
\bmdefine\vj{j}
\bmdefine\vk{k}
\bmdefine\vn{n}
\bmdefine\vp{p}
\bmdefine\vq{q}
\bmdefine\vr{r}
\bmdefine\vs{s}
\bmdefine\vt{t}
\bmdefine\vv{v}
\bmdefine\vw{w}
\bmdefine\vx{x}
\bmdefine\vy{y}
\bmdefine\vz{z}
\bmdefine\vA{A}
\bmdefine\vB{B}
\bmdefine\vC{C}
\bmdefine\vD{D}
\bmdefine\vE{E}
\bmdefine\vF{F}
\bmdefine\vH{H}
\bmdefine\vI{I}
\bmdefine\vJ{J}
\bmdefine\vK{K}
\bmdefine\vL{L}
\bmdefine\vM{M}
\bmdefine\vO{O}
\bmdefine\vP{P}
\bmdefine\vR{R}
\bmdefine\hvR{\hat{R}}
\bmdefine\vS{S}
\bmdefine\vT{T}
\bmdefine\vW{W}
\bmdefine\vX{X}
\bmdefine\vV{V}
\bmdefine\vN{N}
\bmdefine{\vnull}{0}
\bmdefine\valpha{{\alpha}}
\bmdefine\vbeta{{\beta}}
\bmdefine\vgamma{{\gamma}}
\bmdefine\vmu{{\mu}}
\bmdefine\vnu{{\nu}}
\bmdefine\vpi{{\pi}}
\bmdefine\vtau{{\tau}}
\bmdefine\vPi{{\Pi}}
\bmdefine\vomega{{\omega}}
\bmdefine\veta{{\eta}}
\bmdefine\vOmega{{\Omega}}
\bmdefine\vsigma{{\sigma}}
\bmdefine\vSigma{{\Sigma}}
\bmdefine\vTheta{{\Theta}}
\bmdefine\vvarphi{{\varphi}}
\bmdefine\vconst{\mathrm{const}}
\bmdefine\vnabla{{\nabla}}
\bmdefine\vcA{{\mathcal{A}}}
\bmdefine\vcB{{\mathcal{B}}}
\bmdefine\vcL{{\mathcal{L}}}

\DeclareMathOperator{\Det}{Det}

\newcommand{\cA}{\mathcal{A}}
\newcommand{\cB}{\mathcal{B}}

\newcommand{\cD}{\mathcal{D}}

\newcommand{\cK}{\mathcal{K}}

\newcommand{\cM}{\mathcal{M}}

\newcommand{\cR}{\mathcal{R}}

\bmdefine\vcP{{\mathcal{P}}}

\newcommand{\bpm}{\begin{pmatrix}}
	\newcommand{\epm}{\end{pmatrix}}
\newcommand{\bvm}{\begin{vmatrix}}
	\newcommand{\evm}{\end{vmatrix}}

\newcommand*{\varpm}{\mathbin{\smash{\raise-0.3ex\hbox{$\scriptscriptstyle($}}{\pm} \smash{\raise-0.3ex\hbox{$\scriptscriptstyle)$}}}}

\DeclareSymbolFont{operatorsbold}{OT1}{cmr}{bx}{n}
\DeclareMathAccent{\boldbar}{\mathalpha}{operatorsbold}{"16} 
\DeclareFontFamily{OT1}{pzc}{}
\DeclareFontShape{OT1}{pzc}{m}{it}{<-> s * [1.30] pzcmi7t}{}
\DeclareMathAlphabet{\mathpzc}{OT1}{pzc}{m}{it}

\journal{Physics Letters B}

\begin{document}

\begin{frontmatter}

\title{Resolution of Gauss' law in the maximal Abelian gauge}

\author[first]{D.~R.~Junior}
\author[second]{H.~Reinhardt}

\address[first]{Instituto de F\'isica, Universidade Federal Fluminense,
24210-346 Niter\'oi, RJ, Brazil}

\address[second]{Institut f\"ur Theoretische Physik,
Universit\"at T\"ubingen,
Auf der Morgenstelle 14, 72076 T\"ubingen, Germany}
\begin{abstract}
We resolve Gauss' law in the maximal Abelian gauge supplemented by the Coulomb gauge for the Abelian gauge field and derive the gauge-fixed Hamiltonian of QCD.
\end{abstract}

\begin{keyword}

Hamiltonian approach \sep Yang--Mills theory

\end{keyword}

\end{frontmatter}

\section{Introduction}

The maximal Abelian gauge (MAG) accompanied by the Abelian projection was introduced by 't Hooft \cite{r1} and has proved an efficient way to extract the dominant infrared degrees of freedom of Yang-Mills theory, which in this setting emerge as magnetic monopoles. The MAG is the most prominent form of the class of Abelian gauges. In these gauges the coset G/H of the gauge group G is fixed while the Cartan subgroup H is left invariant. Magnetic monopoles arise in these gauges as obstructions to the gauge fixing. Although being gauge fixing artifacts the magnetic monopoles carry the complete topological charge \cite{r2}, \cite{r3}, and by the Atiyah-Singer index theorem  \cite{r4} and the Banks Casher relation \cite{r5} cause spontaneous breaking of chiral symmetry. Lattice calculations also indicate that these monopoles are in fact condensed \cite{r7}, exhibiting the QCD vacuum as a dual superconductor. 

Lattice calculations carried out in the MAG \cite{r6}, \cite{r7} show that the string tension obtained after Abelian projection matches the full string tension within error bars.\footnote{The precise value of the string tension obtained after Abelian projection depends on the quality of gauge fixing achieved.} Furthermore the string tension obtained after Abelian projection is almost entirely produced by the magnetic monopoles \cite{r7}. It should be noted, however, that a monopole Coulomb gas fails to reproduce the difference-of-areas law for double-winding Wilson loops \cite{d-wl}. Thus, although Abelian-projected ensembles correctly reproduce confinement with the expected $N$-ality dependence \cite{nality-2}, the underlying monopole ensemble cannot simply be described as a Coulomb gas. This situation was clarified in the indirect maximal center gauge \cite{rr6}, which consists of first fixing the MAG, then performing Abelian projection, and then fixing to the maximal center gauge. In this Abelian setting the magnetic monopole loops make the center vortex surface non-oriented, a necessary condition to carry non-zero topological charge \cite{r11}.  In this respect, lattice simulations in $SU(2)$ show that almost all (97\%) monopoles and antimonopoles lie on vortex sheets, forming chains \cite{collimation}.
Removing center vortices from the ensemble of gauge field configurations on the lattice the string tension vanishes and chiral symmetry is restored \cite{r12}. Like in the MAG followed by Abelian projection in the maximal center gauge followed by center projection one obtains almost the full string tension \cite{rr6}, \cite{rr7}.


Motivated by these results in ref. \cite{rr12} a wave functional for the Yang-Mills ground state was proposed, in which the gauge field has support exclusively on center vortex configurations. This wave functional produces a perimeter law for the 't Hooft loop \cite{rr13} as well as an area law for the Wilson loop with a string tension which shows Casimir scaling \cite{rr12}. As the center vortex gauge potential (can be and) was chosen Abelian this wave functional should be interpreted in an Abelian projection setting following the MAG. Hence for sophisticated studies of Yang-Mills theory with this wave functional we need the gauge-fixed Hamiltonian in the MAG. Furthermore the Hamiltonian in the MAG is also of interest by itself  since it could be the starting point of an variational calculation analogously to the successful variational calculations carried out in Coulomb gauge \cite{rr5}. A variational calculation in the MAG followed by the Abelian projection had the advantage that one had to deal only with an Abelian gauge field and could hence be simpler. To carry out such calculations the gauge-fixed Hamiltonian of the MAG is required. In the present paper we will show that Gauss' law can be explicitly resolved in this gauge and derive the corresponding gauge-fixed QCD Hamiltonian.

The organization of the paper is as follows:
In the next section we fix our notation, define our gauge fixing, and calculate the corresponding Faddeev-Popov determinant. In sect. 3 we resolve Gauss' law, and derive the gauge-fixed Hamiltonian in sect. 4. In sect. 5 we investigate the Hamiltonian obtained in the Abelian projection. A short summary and some concluding remarks are given in sect.6.

\section{The Maximal Abelian Gauge}
We use anti-hermitian generators $T_A$ of  the gauge group $SU(N)$ which satisfy the algera
\begin{equation}
	[T_A,T_B]=f_{ABC}T_C
\end{equation}
and  are normalized to
\begin{align}
	{\rm Tr}(T_A T_B) = -\frac{1}{2}\delta_{AB}\; .
\end{align} 
Here capital indices $A,B, C,...$ are used for all group generators, i.e. $A=1,\dots ,N^2-1$, 
while lower case indices $a,b,c,...$ and the indices $r,s,t, ...$, respectively, are used for the coset generators $T_a\in SU(N)/U(1)^{N-!} $ and the Cartan generators $T_r$, $r=1,2,\dots ,N-1$, respectively.

The starting point of the Hamiltonian approach to gauge theory is the Weyl gauge
\begin{equation}
	\label{g1}
	A_0(x)=0,
\end{equation}
which is needed for a canonical quantization of the theory since the momentum canonically conjugate to the field coordinate $A_0$ vanishes. The Weyl gauge leaves still invariance w.r.t. time-independent gauge transformations which we fix by first imposing the MAG. For this purpose we split the gauge field $A$ into an Abelian part $\cA$ and a coset part $Q$
\be
\label{g2}
A=\cA +Q, \quad A=A^A T_A, \quad \cA=\cA^r T_r, \quad Q=Q^a T_a.
\ee
The Maximal Abelian gauge \cite{r1} is defined by 
\begin{equation}
	\label{MAG}
	[\cD_k,Q_k]=0, \qquad \cD=\partial+\cA.
\end{equation}
It fixes the gauge freedom associated with coset transformations, but 
leaves invariance 
under Abelian gauge transformations. For the resolution of Gauss' law we need, however, a complete gauge fixing. For this purpose we require in addition to the MAG the Coulomb Gauge for the Abelian part of the gauge field
\begin{equation}
	\label{g3}
	\partial \cdot  \cA=0,
\end{equation}
where the dot $\cdot $ denotes the scalar product in $\mathbb{R}^3$. 
For subsequent calculations it will be convenient to write the gauge conditions in component form. For this purpose we introduce the generators and the gauge field in the adjoint representation.
\begin{equation}
	\label{g4}
	\hat{T}_A^{BC}=f_{BAC}, \qquad \hat{A}^{BC}=A^A\hat{T}_A^{BC}\;.
\end{equation}
In this representation the MAG condition \eqref{MAG} reads
\begin{equation}
	\label{g5}
	\hat{\cD}^{ab}\cdot Q^b=0,  \qquad \hat{\cD}^{ab}= \delta^{ab}\, \partial+\hat{\cA}^{ab}.
\end{equation}
Since $\hat{\cA}^{rs}=\hat{\cA}^{ra}=\hat{\cA}^{ar}=0$ we can compose the two gauge conditions \eqref{g3}, \eqref{g5} into a single equation
\begin{equation}
	\label{g6}
	f^A[A]=\hat{\cD}^{AB}\cdot A^B=0.
\end{equation}
Under a gauge transformation
\begin{equation}
	U=\exp\omega,\qquad \omega= \omega^AT_A
\end{equation}
the gauge field transforms as
\begin{equation}
	A \to A^U=UDU^{-1}.
\end{equation}
For an infinitesimal gauge transformation
\begin{equation}
	\delta U=\exp\delta \omega= 1+\delta \omega+\cdots ,\; \delta \omega= \delta \omega^AT_A
\end{equation}
the gauge field changes by
\begin{equation}
	\label{g7}
	\delta A\equiv	 A^{\delta U}=-\hat{D}\delta \omega= 	-(\hat{\cD}+\hat{Q})\delta \omega,
\end{equation}
which implies for the Abelian gauge field in the adjoint representation the change
\begin{align}
	\delta \hat{\cA}^{AB}=\hat{T}_r^{AB}\delta \cA^r =\hat{T}_r^{AB}(-\hat{D}^{rC})\delta \omega^C.
\end{align}
With this result the gauge fixing functional \eqref{g6} changes under an infinitesimal gauge transformation by
\begin{align}
	\label{g8}
	\delta f^A[A]\equiv f^A[\delta A]&=\hat{\cD}^{AB}\delta A^B+\delta \hat{\cA}^{AB}A^B
	\nonumber \\
	&=[-\hat{\cD}^{AB}\hat{D}^{BC}+\hat{Q}^{Ar}\hat{D}^{rC}]\delta \omega^C,
\end{align}
where we have used the definition \eqref{g4} of $\hat{A}$ and $\hat{\cA}^{Ar}=0$.
From here we obtain for the Faddeev-Popov kernel
\begin{equation}
	\label{g9}
	\cM^{AC}(\vx,\vy)=\frac{\delta f^A[A](\vx)}{\delta \omega^C(\vy)}
\end{equation}
the following result
\begin{equation}
	\label{g10}
	\cM^{AB}(\vx,\vy)=[-(\hat{\cD}^{AC}\hat{\cD}^{CB}+\hat{\cD}^{AC}\hat{Q}^{CB})
	+\hat{Q}^{Ar}\hat{Q}^{rB}]\delta(\vx,\vy).
\end{equation}
In the Cartan subgroup this expression reduces to
\begin{equation}
	\label{g11}
	\cM^{rs}(\vx,\vy)=[-\delta^{rs}\Delta]\delta(\vx,\vy),
\end{equation}
which is the Faddeev-Popov kernel for an Abelian gauge field in Coulomb gauge. From eq. \eqref{g10} one finds in the coset space
\begin{equation}
	\label{g12}
	\cM^{ab}(\vx,\vy)=[-(\hat{\cD}^{ac}\cdot\hat{\cD}^{cb}+\hat{\cD}^{ac}\cdot\hat{Q}^{cb})
	+\hat{Q}^{ar}\cdot\hat{Q}^{rb}]\delta(\vx,\vy),
\end{equation}
and the cross terms
\begin{align}
	\label{g13}
	\cM^{rb}(\vx,\vy) & =[
	-\partial \cdot\hat{Q}^{rb}]\delta(\vx,\vy)\;,
	\nonumber \\
	\cM^{as}(\vx,\vy) &=[-\hat{\cD}^{ac}\cdot\hat{Q}^{cs}	+\hat{Q}^{as}\cdot\partial]\delta(\vx,\vy).
\end{align}
With the Faddeev-Popov determinant at hand we can now proceed to resolve Gauss' law and derive the gauge-fixed Hamiltonian.

\section{Resolution of Gauss' law}

In the Hamilton approach Gauss' law  is a constraint to the wave functional $\Psi(A)\equiv\Psi(\cA, Q)$
\begin{equation}
	\label{g14}
	\hat{D}^{AB}_k\Pi^B_k\Psi(A)=\rho^A\Psi(A)
\end{equation}
which in the absence of external color charges $\rho^A=0$ ensures its gauge invariance. Here 
\begin{equation}
	\label{g15}
	\hat{D}=\partial+\hat{A}=\hat{\cD}+\hat{Q}
\end{equation}
is the covariant derivative in the adjoint representation \eqref{g4} and $\Pi$ is the momentum operator of the total gauge field $A^A_k $, which in the coordinate representation reads:
\begin{equation}
	\label{g16}
	\Pi^A_k(x)=\frac{\delta}{i\delta A^A_k(x)}.
\end{equation}
Analogously to the gauge field \eqref{g2} we split the momentum operator into its Abelian and non-Abelian parts:
\be 
\label{g17}
\Pi=\pi +P, \quad \Pi=\Pi^A T_A, \quad \pi=\pi^r T_r, \quad P=P^a T_a
\ee
which in the coordinate representation \eqref{g16} are given by
\begin{equation}
	\label{g17}
	\pi^r_k(x)=\frac{\delta}{i\delta \cA^r_k(x)}, \qquad P^a_k(x)=\frac{\delta}{i\delta Q^a_k(x)}.
\end{equation}
Since the MAG distinguishes between the Abelian and non-Abelian components of the gauge field we also split Gauss' law into its Cartan and coset parts:
\begin{equation}
	\label{g19}
\partial \cdot  \Pi^r+\hat{Q}^{rb} \cdot  \Pi^b	=\rho^r,
\end{equation}
\begin{equation}
	\label{g20}
		\hat{\cD}^{ab}\cdot \Pi^b+\hat{Q}^{aB}\cdot \Pi^B=\rho^a,
\end{equation}
where we have used that
\begin{equation}
	\label{g21}
\hat{\cA}^{rA}=0, \quad 	
\hat{\cA}^{Ar}=0, \quad \hat{Q}^{rs}=0,
\end{equation}
which is a consequence of  $f_{rsA}=0$. This also implies:
\begin{equation}
	\label{g22}
	\hat{\cD}^{rs}=\delta^{rs} \partial, \quad	\hat{\cD}^{ar}=0, \quad  \hat{\cD}^{ra}=0.
\end{equation}
In eqs. \eqref{g19},  \eqref{g20} we have omitted the wave functional $\Psi{(\cA,Q)}$ in order to prevent the subsequent equations from getting cluttered. But we emphasize that these equations  and the ones following from them are not operator identities but constraints to the wave functional $\Psi(\cA, Q)$.

To resolve Gauss' law in the MAG we proceed analogously to its resolution in Coulomb  gauge by introducing the (generalized) longitudinal and transversal projectors:
\begin{equation}
		\label{g23}
	l_{ij}=\partial_i (-\Delta)^{-1}\partial_j, \quad t_{ij}=\delta_{ij}-l_{ij},
\end{equation}
\begin{equation}
		\label{g24}
	L_{ij}^{ab}=\hat{\cD}^{ac}_i[(-\hat{\cD}_m\hat{\cD}_m)^{-1}]^{cd}\hat{\cD}^{db}_j, \quad T_{ij}^{ab}=\delta_{ij}\delta^{ab} -L_{ij}^{ab}
\end{equation}
for the Abelian and coset fields, respectively. With these projectors we split these fields and their momenta into their longitudinal and transverse parts:
\begin{equation}
		\label{g25}
	\cA=\cA^{||}+\cA^{\bot}, \quad Q=Q^{||}+Q^{\bot}, 
\end{equation}
\begin{equation}
		\label{g26}
	\pi=\pi^{||}+\pi^{\bot}, \quad P=P^{||}+P^{\bot}, 
\end{equation}
where
\begin{equation}
		\label{g27}
	\cA^{||r}_i=l_{ij}\cA^r_j, \quad \cA^{\bot r}_i=t_{ij}\cA^r_j,
\end{equation}
\begin{equation}
		\label{g28}
	Q^{||a}_i=L_{ij}^{ab}Q^b_j, \quad Q^{\bot a}_i=T_{ij}^{ab}Q^b_j
\end{equation}
and analogous equations for the momentum operators.
Note that in terms of the projected fields the gauge conditions \eqref{g3}, \eqref{MAG} read:
\begin{equation}
	\label{g29}
	A^{||}=0, \quad 	i.e.  \quad \cA^{||}=0, \quad Q^{||}=0.
\end{equation}
Imposing these gauge conditions we are left with the transverse gauge fields only. We will hence skip the subscript $\bot$ at the fields $\cA$, $Q$ in the following. The vanishing of the longitudinal gauge fields does, however, not imply that the longitudinal momentum operators $\Pi^{||}=(\pi^{||}, P^{||})$ also vanish although they can no longer be expressed by the variational derivatives \eqref{g17} w.r.t. the longitudinal gauge fields  $A^{||}=(\cA^{||}, Q^{||})$. Instead the longitudinal momentum operators are completely determined by Gauss' law as we will show in the following. \footnote{Note that Gauss' law cannot be explicitly resolved in any arbitrary gauge. Since the MAG is a non-linear gauge it is by no means obvious that Gauss' law can be resolved in this gauge.}

After the splitting of the momentum operators into longitudinal and transverse parts and using
\begin{equation}
	\partial \cdot \pi^{\bot}=0, \qquad \hat{\cD}\cdot  P^{\bot}=0
\end{equation}
Gauss' law \eqref{g19}, \eqref{g20} becomes
\begin{equation}
	\label{g30}
	\partial \cdot \pi^{||r}	=\rho^r-\hat{Q}^{rb} \cdot  P^{||b}-\hat{Q}^{rb} \cdot P^{\bot b}
\end{equation}
\begin{equation}
		\label{g31}
	(\hat{\cD}^{ab}+\hat{Q}^{ab})\cdot  P^{||b}+	\hat{Q}^{ar} \cdot  \pi^{||r}=\rho^a -\hat{Q}^{aB} \cdot  \Pi^{\bot B}.
\end{equation}
The first equation is exactly resolved as in Coulomb gauge yielding
\begin{align}
	\label{g32}
	 \pi^{||r}_k
	=-\partial_k(-\Delta)^{-1}[\rho^r-\hat{Q}^{rb}\cdot P^{||b}-	\hat{Q}^{rb}\cdot P^{\bot r}	].
\end{align}
Inserting this result into the non-Abelian component \eqref{g20} of Gauss' law we find the relation:
\begin{align}
	\label{g33}
(	\hat{\cD}^{ab}+	\hat{Q}^{ab}+\hat{Q}^{ar}\cdot\partial(-\Delta)^{-1}	\hat{Q}^{rb})\cdot P^{||b}
 =\rho^a+\rho^a_{YM} ,
\end{align}
where we have defined the non-Abelian Yang-Mills charge in MAG:
\begin{equation}
		\label{g34}
	\rho^a_{YM}= -\hat{Q}^{aB}\cdot \Pi^{\bot B}+ \hat{Q}^{ar} \cdot \partial (- \Delta)^{-1}[\rho^r-\hat{Q}^{rb}\cdot P^{\bot b}].
\end{equation}
Note that
\begin{equation}
	\hat{Q}^{aB}\cdot \Pi^{\bot B}=\hat{Q}^{as}\cdot \pi^{\bot s}+\hat{Q}^{ab}\cdot P^{\bot b}.
\end{equation}
The longitudinal component of the non-Abelian momentum operator can be represented as:
\begin{equation}
		\label{g35}
	P^{||a}_k=-	\hat{\cD}^{ab}_k \chi^b,
\end{equation}
where $\chi^a(x)$ is a (Lorentz-)scalar operator living in the color coset space. With this ansatz the non-Abelian Gauss law \eqref{g33} becomes
\begin{align}
		\label{g36}
	\int d^3y \cK^{ab}(\vx,\vy)\chi^{b}(\vy)
	=\rho^a(\vx)+\rho^a_{YM}(\vx),
\end{align}
where we have defined the integral kernel
\begin{align}
		\label{g37}
	\cK^{ab}(\vx,\vy) &=	-\big[	\hat{\cD}^{ac}_k\hat{\cD}^{cb}_k+	\hat{Q}^{ac}_k\hat{\cD}^{cb}_k
	\nonumber \\
	&	+\hat{Q}^{ar}_l\partial_l(-\Delta)^{-1}	\hat{Q}^{rc}_k\hat{\cD}^{cb}_k\big]\delta(\vx,\vy).
\end{align}
Equation \eqref{g36} is solved for
\begin{equation}
		\label{g38}
	\chi^a= (\cK^{-1})^{ab}(\rho^b+\rho^b_{YM}).
\end{equation}
The longitudinal non-Abelian momentum operator of the coset field is then given by
\begin{equation}
		\label{g39}
	P^{||a}_k=-	\hat{\cD}^{ab}_k  (\cK^{-1})^{bc}(\rho^c+\rho ^c_{YM}).
\end{equation}
This expression has to be inserted into eq \eqref{g32} to obtain the final representation of the longitudinal Abelian part of the momentum operator of the gauge field, which yields
\begin{align}
		\label{g40}
	\pi^{||r}_k&=-\partial_k(-\Delta)^{-1}[\rho^r+\cR^r_{YM}].
\end{align}
Here we have defined the Abelian Yang-Mills charge in MAG
\begin{equation}
		\label{g41}
	\cR^r_{YM}=-\hat{Q}^{rb}_kP^{\bot b}_k+\hat{Q}^{rb}_k \hat{\cD}^{bc}_k  (\cK^{-1})^{cd}(\rho^d+\rho^d_{YM}).
\end{equation} 
Note that above derived expressions \eqref{g39}, \eqref{g40}
for the longitudinal momentum operators are not operator identities but are valid only when acting on the wave functional $\Psi(A)\equiv \Psi(\cA. Q)$, see eq. \eqref{g14}.

\section{The Gauge-Fixed Hamiltonian}

With the above expressions for the longitudinal momentum operators obtained from the resolution of Gauss' law we can derive the gauge-fixed Hamiltonian proceeding analogously as in the case of Coulomb gauge. For this purpose we start from the kinetic energy of the Yang-Mills field before gauge fixing, 
\begin{align}
	E_k=\int \cD A \Psi^*(A)\left[\frac{1}{2}\int d^3 x \Pi(\vx)\cdot \Pi(\vx)\right] \Psi(A),
\end{align} 
and split the momentum operator by means of the projectors eqs. \eqref{g23}, \eqref{g24} into its longitudinal and transverse parts. (Note these projectors can be used independently of (and hence before) the gauge fixing.) Using $\Pi^{||} \cdot \Pi^\bot=0$ we have
\begin{align}
E_k&
=\int \cD A \Psi^*(A)\left[\frac{1}{2}\int d^3 x [\Pi^{||}(\vx)\cdot \Pi^{||}(\vx)+\Pi^\bot(\vx)\cdot \Pi^\bot(\vx)] \right]\nonumber\\&\Psi(A).
\end{align} 
Before gauge-fixing the representation \eqref{g16} of the momentum operators is valid, i.e.
\begin{equation}
	\label{g16a}
	\Pi^{||}(x)=\frac{\delta}{i\delta A^{||}(x)}, \qquad \Pi^{\bot}(x)=\frac{\delta}{i\delta A^{\bot}(x)}.
\end{equation}
We can use this representation to perform a partial functional integration yielding
\begin{align}
	\label{g42}
	E_k=\int \cD A \frac{1}{2}\int d^3 x \Big[&(\Pi^{||}(\vx) \Psi(A)) ^* \cdot (\Pi^{||}(\vx) \Psi(A))
	\nonumber \\
	&+(\Pi^{\bot}(\vx) \Psi(A)) ^* \cdot (\Pi^{\bot}(\vx) \Psi(A))   \Big].
\end{align} 
We now fix the gauge \eqref{g29}, $A^{||}=0$, by means of the Faddeev-Popov method obtaining
\begin{align}
		\label{g45}
	E_k=&\int \cD A^\bot J(A^\bot) \frac{1}{2}\int d^3 x \Big[(\Pi^{||}(\vx) \Psi(A^\bot))^* \cdot (\Pi^{||}(\vx) \Psi(A^\bot))
	\nonumber \\
	&+(\Pi^{\bot}(\vx) \Psi(A^\bot)) ^* \cdot (\Pi^{\bot}(\vx) \Psi(A^\bot))   \Big],
\end{align} 
where
	\label{g46}
\begin{equation}
	J(A^\bot)= {Det}\cM
\end{equation}
is the Faddeev-Popov determinant with the kernel $\cM$ defined in eq. \eqref{g10}. The transversal momentum operators $\Pi^{\bot}$ are still given by the functional derivative  \eqref{g16a} w.r.t. the transverse gauge fields $A^{\bot}$, 
while the longitudinal momentum operators $\Pi^{||}=(\pi^{||}, P^{||})$ are given by the expressions \eqref{g39}, \eqref{g40} derived above from Gauss' law. These expression are entirely given by the transverse gauge fields $A^\bot=(\cA^\bot, Q^\bot)$, the transverse momenta $\Pi^{\bot}=(\pi^\bot, P^\bot)$ and the color density of the matter fields $\rho^A$.  Note also that in eq. \eqref{g45} we do not need the longitudinal momentum operator $\Pi^{||}$ itself but only its action on the wave functional, which indeed is completely determined by Gauss' law.

Performing in eq.\eqref{g45} a partial functional integration in the variables $A^{\bot}=( \cA^{\bot}, Q^{\bot})$ we arrive at\footnote{Note that the longitudinal momenta $\Pi^{||}=(\pi^{||},P^{||})$ are linear functionals of the transverse momenta  $\Pi^{\bot}=(\pi^{\bot}, P^{\bot})$, see eqs. \eqref{g39}, \eqref{g34}, \eqref{g40} and \eqref{g41}.}
\begin{align}
	\label{g48}
	E_k=&\int \cD A^\bot J(A^\bot)  \Psi^*(A^\bot)\frac{1}{2}\int d^3 x \Big[J^{-1}(A^\bot)\Pi^{||}(\vx) \cdot J(A^\bot) \Pi^{||}(\vx) 
	\nonumber \\
	&+J^{-1}(A^\bot) \Pi^{\bot}(\vx)  \cdot  J(A^\bot) \Pi^{\bot}(\vx)    \Big]\Psi(A^\bot)\;.
\end{align}  
Since this expression is valid for any wave functional $\Psi(A^\bot)$ we can read from it the kinetic part of the gauge-fixed Hamiltonian $H_k^{gf}$ by identifying the above expression \eqref{g48} with
\begin{equation}
	E_k=\int \cD A^\bot J(A^\bot)  \Psi^*(A^\bot)  H_k^{gf} \Psi(A^\bot)
\end{equation}
yielding
\begin{align}
	\label{g49}
	 H_k^{gf}=\frac{1}{2}\int d^3 x \Big[&J^{-1}(A^\bot)\Pi^{||}(\vx) \cdot J(A^\bot) \Pi^{||}(\vx) 
	\nonumber \\
	&+J^{-1}(A^\bot) \Pi^{\bot}(\vx)  \cdot  J(A^\bot) \Pi^{\bot}(\vx)    \Big].
\end{align}
Adding the potential part, which is given by the magnetic energy, we obtain the gauge-fixed Hamiltonian of Yang-Mills theory in the gauge \eqref{g29}
\begin{align}
	\label{g50}
	H^{gf}=&\frac{1}{2}\int d^3 x \Big[J^{-1}(A^\bot)\Pi^{||}(\vx) \cdot J(A^\bot) \Pi^{||}(\vx) 
	\nonumber \\
	&+J^{-1}(A^\bot) \Pi^{\bot}(\vx)  \cdot  J(A^\bot) \Pi^{\bot}(\vx)  +B(A^\bot)  \cdot B(A^\bot)  \Big]\;.
\end{align}
Here $ B(A^\bot)$ is the non-Abelian magnetic field of the transverse gauge field $A^\bot$. Furthermore the transversal momentum operator $\Pi^{\bot}$  is still given by eq. \eqref{g16a} while the longitudinal momentum operator $\Pi^{||}=(\pi^{||}, P^{||})$ is given now by eqs. \eqref{g39} and \eqref{g40}. Due to the kinetic part of the longitudinal momenta this Hamiltonian is quite involved. Fortunately the lattice results obtained in the MAG \cite{r7}, \cite{r6} indicate that the infrared and, in particular, the confining properties of Yang-Mills theory can be 
extracted from the Abelian projected theory. Therefore we will now implement the Abelian projection in the gauge-fixed Hamiltonian \eqref{g50}.

\section{The Abelian projected theory}

In the Abelian projection the coset field is put to zero. With $Q=0$ the Yang-Mills charges \eqref{g34} vanish, $\rho^A_{YM}=0$, and the integral kernel $\cK$ \eqref{g37} reduces to
\begin{align}
	\label{g55}
	\cK^{ab}(\vx,\vy) &=-\hat{\cD}^{ac}_k\hat{\cD}^{cb}_k\delta(\vx,\vy)=\cM^{ab}(\vx,\vy)
\end{align}
and equals then the coset part of the Faddeev-Popov kernel $\cM^{ab}(\vx,\vy)$ \eqref{g10}. Furthermore the gauge-fixed Hamiltonian \eqref{g50} reduces to 
\begin{align}
	\label{g56}
	&H^{gf}=\frac{1}{2}\int d^3 x \Big[
	J^{-1}(\cA) \pi(\vx)  \cdot  J(\cA) \pi(\vx)  +B(\cA)  \cdot B(\cA)\Big]+
	\nonumber \\
	& \frac{1}{2}\!\int\! d^3x\!\!\int\! d^3y\! \left[\rho^r(\vx)(-\Delta)^{-1} \delta (\vx,\vy)\rho^r(\vy) +\rho^a(\vx)\left(\cM^{-1}\right)^{ab} \rho^b(\vx)\right],
\end{align}
where $\pi(\vx)=\delta/i \delta \cA(\vx)$ is the momentum operator and $\cB=\partial \times \cA$ the magnetic field of the transverse Abelian gauge field $\cA \equiv \cA^\bot$. Note that the matter color charge $\rho^A$ enters this Hamiltonian asymmetrically. While the non-Abelian components $\rho^a$ experience a nontrivial interaction mediated by the kernel $\cM^{-1}$ \eqref{g55} the Abelian components $\rho^r$ interact via the ordinary Coulomb potential. Does this mean that the Abelian color charges are not confined? Certainly not. The resolution of this puzzle comes when one adds the Hamiltonian of the matter fields, which for the quarks is given by
\begin{align}
	\label{g57}
	H^{gf}_q=\int d^3 x  \left[\bar{q}(\vx)\alpha \cdot \vp q(\vx) +\alpha \cdot\cA^r(\vx) \rho^r(\vx) \right]
\end{align}
with $\rho^r(\vx)= \bar{q}(\vx) iT^r q(\vx)$ being the Abelian color charge of the quarks. Only this color charge couples to the gauge field $\cA$. Although being Abelian this gauge field has a nontrivial kinetic Hamiltonian \eqref{g56} and hence a nontrivial propagator. Therefore its exchange between the Abelian charges produces a nontrivial potential for the latter.

It is interesting and illuminating to compare the gauge-fixed Hamiltonian \eqref{g56} in the MAG and Abelian projection with the Hamiltonian in Coulomb gauge \cite{rr5}:
\begin{align}
	\label{g58}
	H^{gf}=&\frac{1}{2}\int d^3 x \Big[
	J^{-1}(A^\bot) \Pi^{\bot}(\vx)  \cdot  J(A^\bot) \Pi^{\bot}(\vx)  +B(A^\bot)  \cdot B(A^\bot)  \Big]\nonumber\\&+H_C.
\end{align}
Here the Faddeev-Popov determinant is given by
\begin{equation}
	\label{g60}
	J(A^\bot)=\Det(-\hat{D}\cdot \partial)
\end{equation}
and
\begin{align}
	\label{g59}
	H_C&=\frac{1}{2}\int d^3 x J^{-1}(A^\bot)\Pi^{||}(\vx) \cdot J(A^\bot) \Pi^{||}(\vx) 
	\nonumber \\
	&= \frac{1}{2}\int d^3x \int d^3y  \rho^A_{tot}(\vx)F^{AB}(\vx,\vy) \rho^B_{tot}(\vx)
\end{align}
is the so-called Coulomb term, which arises from the longitudinal part of the kinetic energy, $\int \Pi^{||} \cdot \Pi^{||} $. It describes a nontrivial interaction of the total color charge
\begin{equation}
	\rho_{tot}^A=\rho^A+\rho_{YM}^A,
\end{equation}
which contains besides the color charge of the matter fields, $\rho^A$, also a color charge of the gauge field
\begin{equation}
	\rho_{YM}^A=-\hat{A}^{AB} \cdot \Pi^B\;.
\end{equation}
Such a charge is missing in the Abelian projected Hamiltonian \eqref{g56} (but is, of course, present in the Hamiltonian in the MAG before the Abelian projection, see eq.\eqref{g50}). Furthermore, contrary to the latter, the Hamiltonian in Coulomb gauge contains the full non-Abelian magnetic energy.

The variational calculation carried out in Coulomb gauge in ref. \cite{rr5} shows that for the confining properties of the gluon propagator neither the Coulomb term $H_C$ \eqref{g59} nor the non-Abelian part of the magnetic energy is relevant. The infrared behaviour of the gluon propagator is exclusively determined by the nontrivial kinetic term in the Hamiltonian, i.e. by the Faddeev-Popov determinant in the kinetic energy. The kinetic term in the Abelian projected Hamiltonian \eqref{g56} has the same structure as the one in Coulomb gauge \eqref{g58} and the Faddeev-Popov determinant in the MAG after Abelian projection \eqref{g55} is quite similar to the one Coulomb gauge \eqref{g55}. We can therefore expect that the Hamiltonian in MAG after Abelian projection \eqref{g56} yields a confining gluon propagator.

\section{Summary and Conclusions}

In this paper we have resolved Gauss' law in the MAG supplemented by the Coulomb gauge for the Cartan components of the gauge field and derived the corresponding gauge-fixed Hamiltonian. This Hamiltonian is much more complicated than e.g. the one in Coulomb gauge. However, lattice calculations show that in the MAG one captures the confining properties even after Abelian projection. In the Abelian projection the gauge-fixed Hamiltonian simplifies tremendously and becomes even simpler than the one in Coulomb gauge. As we have argued by the findings in the variational calculations in Coulomb gauge the Hamiltonian obtained in the MAG after Abelian projection contains nevertheless the essential non-Abelian features to describe the confining properties of Yang-Mills theory.
It would be interesting to determine the vacuum wave functional of this Hamiltonian by the variational principle and calculate from it the static quark potential.

\end{document}